\documentclass[aps,twocolumn,showpacs,floatfix]{revtex4}
\usepackage{graphicx}% Include figure files
\usepackage{psfig}
\usepackage{epsfig}

\begin{document}
%\draft

%%%%%%%%%%%%%%%%%%%%%
% Macros            %
%%%%%%%%%%%%%%%%%%%%%
\def\kket{\rangle \mskip -3mu \rangle}
\def\bbra{\langle \mskip -3mu \langle}

\def\ket{\rangle}
\def\bra{\langle}

\def\pard{\partial}

\def\sinh{{\rm sinh}}
\def\sgn{{\rm sgn}}

%Characters
\def\alp{\alpha}
\def\del{\delta}
\def\Del{\Delta}
\def\eps{\epsilon}
\def\gam{\gamma}
\def\sig{\sigma}
\def\kap{\kappa}
\def\lam{\lambda}
\def\ome{\omega}
\def\Ome{\Omega}

\def\th{\theta}
\def\vphi{\varphi}

\def\Gam{\Gamma}
\def\Ome{\Omega}

\def\kav{{\bar k}}

\def\abf{{\bf a}}
\def\cbf{{\bf c}}
\def\dbf{{\bf d}}
\def\gbf{{\bf g}}
\def\kbf{{\bf k}}
\def\lbf{{\bf l}}
\def\nbf{{\bf n}}
\def\pbf{{\bf p}}
\def\qbf{{\bf q}}
\def\rbf{{\bf r}}
\def\ubf{{\bf u}}
\def\vbf{{\bf v}}
\def\xbf{{\bf x}}
\def\Cbf{{\bf C}}
\def\Dbf{{\bf D}}
\def\Kbf{{\bf K}}
\def\Pbf{{\bf P}}
\def\Qbf{{\bf Q}}

\def\omet{{\tilde \ome}}
\def\gammat{{\tilde \gamma}}
\def\Ft{{\tilde F}}
\def\It{{\tilde I}}
\def\ut{{\tilde u}}
\def\bt{{\tilde b}}
\def\vt{{\tilde v}}
\def\xt{{\tilde x}}

\def\ph{{\hat p}}

\def\wt{{\tilde w}}
\def\xit{{\tilde \xi}}
\def\phit{{\tilde \phi}}
\def\rhot{{\tilde \rho}}

\def\Cb{{\bar C}}
\def\Nb{{\bar N}}
\def\Ab{{\bar A}}
\def\Db{{\bar D}}
\def\gb{{\bar g}}
\def\nb{{\bar n}}
\def\bb{{\bar b}}
\def\Pib{{\bar \Pi}}
\def\rhob{{\bar \rho}}
\def\phib{{\bar \phi}}
\def\psib{{\bar \psi}}
\def\omeb{{\bar \ome}}

\def\Sh{{\hat S}}
\def\Wh{{\hat W}}

\def\SS{I}
\def\psiw{{\xi}}
\def\tI{{g}}

\def\Ep#1{Eq.\ (\ref{#1})}
\def\Eqs#1{Eqs.\ (\ref{#1})}
\def\EQN#1{\label{#1}}

\newcommand{\beqa}{\begin{eqnarray}}
\newcommand{\eeqa}{\end{eqnarray}}

%%%%%%%%%%%%%%%%%%%%%
% Title and authors %
%%%%%%%%%%%%%%%%%%%%%
\title{Complex collective states  in a one-dimensional  two-atom system}
\author{Gonzalo Ordonez and  Sungyun Kim}
\address{Center for Studies in Statistical Mechanics and Complex Systems,
         \\ The University of Texas at Austin, Austin, TX 78712 USA}
\date{\today}

%%%%%%%%%%%%%%%%%%%%
% Abstract         %
%%%%%%%%%%%%%%%%%%%%

\begin{abstract}
We consider a pair of identical  two-level atoms  interacting with
a scalar field in one dimension, separated by a
distance $x_{21}$. We restrict our attention to states where one atom is excited and the other is in the ground state, in symmetric or anti-symmetric combinations. We obtain exact collective decaying states, belonging to
a complex spectral representation of the Hamiltonian. The imaginary parts of the
eigenvalues give the decay rates, and the real parts give the
average energy  of the collective states. In one dimension there is strong interference between the fields emitted by the atoms, leading to long-range cooperative effects. The decay rates and
the energy oscillate with the distance
$x_{21}$.  Depending on $x_{21}$,  the decay rates will  either decrease, vanish or increase as compared with the one-atom decay rate. We have sub- and super-radiance at periodic intervals.  Our model may be used to study  two-cavity electron wave-guides. The vanishing of the collective decay rates then suggests the possibility of obtaining  stable configurations, where an electron is trapped inside the two cavities.
\end{abstract}


\maketitle

%%%%%%%%%%%%%%%%%%%%
% Section 1 %
%%%%%%%%%%%%%%%%%%%%
\section{Introduction}
\label{sec:Int}

Systems of interacting atoms form collective states where the atoms behave differently from isolated ones \cite{Dicke,Woger}.
For atoms in their ground states the collective effects are relatively small. They produce the van-der Waals or Casimir-Polder forces between atoms \cite{Compagno}, associated with the cloud of virtual photons surrounding the atoms.

When the atoms are in excited states  they  can exchange real photons originated by spontaneous emission.  Depending on the situation, the real photons can give strong collective effects, altering the forces between atoms and also  their rate of spontaneous emission  \cite{Dicke,Stephen, Milonni}.  For example for identical atoms in  three  dimensions separated by small  distances, the decay rate will essentially double or vanish,   depending on whether the initial state  is symmetric or antisymmetric with respect to exchange of atoms \cite{Stephen, Dung}. We have ``super-radiance'' or ``sub-radiance,'' respectively. If the atoms are not identical \cite{Power}, or if the distance between the atoms is larger than their characteristic wavelengths, the effects become much smaller \cite{Stephen}.

Most  studies on two-atoms  (see, e.g., Refs. \cite{Dicke,Woger,Compagno, Stephen, Milonni,Dung,Power,Dung2, Kweon,Ficek}) focus on three-dimensional systems. In this paper we will consider a one-dimensional system. We will show that  the exchange of real photons gives strong collective effects even for large separations between the atoms. 

Our system is analogous to electron wave guides consisting of two cavities connected by  a lead \cite{Suresh0,Suresh}.  Hence the effects we will discuss may be studied experimentally.

We will consider a simplified model where we have two identical two-level atoms, with basis states where   one atom
is excited, while the other  is in the ground state. We will 
use  the dipole and rotating-wave approximations for the interaction with the field.

We will describe  the collective two-atom states through  complex eigenstates of the Hamiltonian  \cite{Nakanishi, Sudarshan, Bohm, PPT}.  The complex eigenstates  decay exponentially in time,  breaking time-symmetry. The real and imaginary parts of the complex eigenvalues give the average  energy of the collective states and the emission rates, respectively.

The Hamiltonian, being a Hermitian operator, can only have complex eigenvalues if the eigenstates do not belong to a  Hilbert space.   In one-atom systems the non-Hilbertian nature of the complex states is manifested in their field intensity,  which includes a factor growing exponentially with the distance from the atom. This growth in space  is related to the exponential decay of the atom in time \cite{POP2001}. The exponential field  inside the light-cone  of the atom represents the real emitted photons.  As we will show, this field has physical effects, which can be seen adding a second atom.

 Outside the light-cone, the field associated with the complex eigenstates grows exponentially  without truncation. However, including the complete set of eigenstates in the complex representation of the Hamiltonian, the exponential field outside the light-cone is cancelled   by 
 renormalized field states  \cite{POP2001}.  This is consistent with causality, because 
 the field further away from the atom is emitted earlier. Going further away, one reaches the point corresponding to the time where the atom was excited.  At this point the field stops growing. 

One can  introduce  complex states that are truncated outside the light-cone, using distributions dependent  on  the test functions or observables. These are considered in Ref. \cite{Sungyun}.

In the two-atom system, the real photons emitted by each atom are absorbed by the other atom. In $d$ dimensions the emitted field includes a $1/r^{d-1}$ decrease factor with distance. Hence the field has a strong effect in $d=1$ dimensions, as compared  with $d=2$ or $3$ dimensions. 

We will show that in one dimension  the decay rates of the collective states oscillate with the distance between the atoms. In contrast to Dicke's states \cite{Dicke}, both symmetric and antisymmetric states of the two atoms can become sub-radiant and super-radiant as the distance between the two atoms is varied. For distances that are integer multiples of the atom wavelength,  the collective decay rates vanish, leading to  stable collective states.  These states can trap field energy between the atoms.

In Sec. \ref{sec:Sum} we briefly discuss the complex representation of the Hamiltonian for a one-atom system. In Sec. \ref{sec:two}  we introduce our two-atom model and its complex collective eigenstates. In Secs.  \ref{sec:emergence} and \ref{sec:b} we discuss the emergence of the collective states and the bouncing of photons between the atoms. In Sec. \ref{sec:x12} we consider the decay rate and average energy of the collective states as a function of the distance between the atoms. We discuss super-radiance and sub-radiance, including stable collective states  mentioned above. We also give a heuristic discussion on the force between the atoms.  In Sec. \ref{wg} we discuss the mapping of our model to a two-cavity electron wave guide, and show that this system allows an approximate stable collective state.

%%%%%%%%%%%%%%%%%%%%
% Section 2 %
%%%%%%%%%%%%%%%%%%%%
\section{One-atom system}
\label{sec:Sum}

In order to introduce the complex spectral representation of the Hamiltonian, we consider first a single two-level atom interacting with a field  in one-dimensional space. This is  the  Friedrichs-Lee
model in one dimension.  We briefly review the
main results. More details can be found in Ref. \cite{PPT}.

\begin{figure}[htb]
\begin{center}
\includegraphics[width=2.5in]{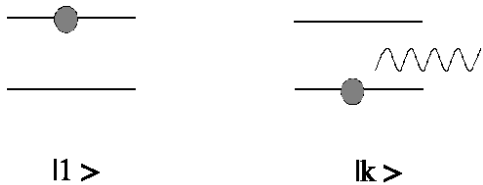}
\caption{One-atom system} \label{1fig}
\end{center}
\end{figure}

 The Hamiltonian is given by 
\begin{eqnarray}
   H &=& {H}_{0}\ + \lambda   V\EQN{(2.1)}\\
     &=& \omega_{1}|1\ket\bra 1| + \sum\limits_{k}\omega_{k}|k\ket \bra k|
     +\lambda\sum\limits_{k} V_{k}  (|k\ket\bra 1| + |1\ket\bra
     k|).\nonumber
\end{eqnarray}
where we put $c=\hbar =1$. The state $|1\ket$ represents the bare
atom in its excited level with no field present, while the state
$|k\ket$ represents a bare field mode (``photon'') of momentum $k$
together with the atom in its ground state (see Figure \ref{1fig}).

The energy of the ground state is chosen to be zero; $\ome_1$  is
the bare energy of the excited level and $\ome_k\equiv |k|$ is the
photon energy. The coupling
constant $\lam \ll 1$ is dimensionless. We assume periodic
boundary conditions. We put the system in a ``box'' of size $L$
and  take the limit $L\to\infty$.  For $L$ finite the
momenta  $k$ are discrete. In the limit $L\to\infty$ they become
continuous, i.e., 
\begin{equation}
{ 2\pi  \over L}  \sum_k \to  \int dk.
 \EQN{sumk}
\end{equation}
We have $\bra a|b\ket = \del_{a,b}=$ Kroenecker delta. In the limit $L\to\infty$,
\begin{equation}
 { L \over 2\pi  } \del_{k,k'} \to  \del(k-k').
 \EQN{delKr}
\end{equation}

The interaction term is obtained through the dipole approximation as well as the rotating-wave approximation.
The potential $V_k$ is of order $L^{-1/2}$. For convenience we write
\begin{equation}
 V_k = (2 \pi/L)^{1/2} v_k,
 \EQN{Veka}
\end{equation}
where $v_k$ is of order $1$ in the continuous spectrum limit $L\to \infty$.
As a specific example we will assume that \cite{Drude,FP99}
\begin{eqnarray}
  v_k  =  v(\ome_k) = { \ome_k^{1/2} \over [1 +(\ome_k/\ome_M)^2]^n}
 \EQN{Vomeka}
\end{eqnarray}
with $n=1$. The constant $\ome_M^{-1}$ determines the
range of the interaction. We shall assume that the
interaction is of short range, i.e., $\ome_M \gg \ome_1$.

The state $|1\ket$ is  unstable if
 \begin{equation}
  \ome_1 > \int_{-\infty}^{\infty} dk\, \frac{\lam^2 v_k^2}{\ome_k}
 \EQN{w1u}
\end{equation}
Otherwise, it is stable  \cite{cohen}. Hereafter we will  consider the unstable
case.

 In the unstable  case one can construct renormalized field eigenstates that diagonalize the Hamiltonian as
 \begin{equation}
  H = \sum_k|\phit_k^\pm\ket \ome_k \bra\phit_k^\pm|
 \EQN{FreH}
\end{equation}
where
 \begin{equation}
  \lim_{\lam \to 0} |\phit_k^\pm\ket = |k\ket
 \EQN{phikk}
\end{equation}
and hereafter we use the summation over field modes in the sense of \Ep{sumk}.
The index $\pm$ refers to either ``in'' or ``out'' scattering eigenstates.

The explicit form of the eigenstates is given by \cite{PPT}
 \begin{equation}
    |\phit_k^\pm\ket    =
   |k\ket +  {\lambda V_k \over \eta^\pm(\omega_k)}
   \left[|1\ket + {\sum_l} {\lambda V_l  \over
   \omega_k-\omega_l\pm i\epsilon}|l\ket\right]
 \EQN{AA6}
\end{equation}
where $\eps$ is an infinitesimal positive number. The limit  $\eps\to 0$ is taken after  the limit $L\to\infty$.
In \Ep{AA6},
\begin{eqnarray}
 \eta^\pm(\ome)  &\equiv&  \ome -\omega_1 - \sum_{k'} {\lambda^2 V_{k'}^2 \over
 (\ome-\ome_{k'})^\pm}\nonumber\\
&=& \ome -\omega_1 - 2 \int_0^\infty dk'\, {\lambda^2 v_{k'}^2 \over
 (\ome-k')^\pm}
 \EQN{(2.10)}
\end{eqnarray}
is the inverse of Green's function.  The $+$ (or $-$) superscript in \Ep{(2.10)}
indicates analytic continuation from the upper (or lower)
half-plane of $\ome$  \cite{PPT}.  Using the complex  delta-function $\del_C$ \cite{Nakanishi}
we can write
\beqa
 \frac{1}{(\ome-k')^\pm} =
 \left\{ \begin{array}{ll}
        (\ome-k')^{-1}, &  \pm {\rm Im\, } \ome >  0\\            
       (\ome-k')^{-1} \mp  2\pi i \del_C(k'-\ome),  &  \pm {\rm Im\, } \ome < 0     
                 \end{array} \right. \nonumber\\
                 \EQN{complexd}
 \eeqa

When $\ome$ is real, we have
  \beqa
  \eta^{\pm}(\ome) \equiv \ome-\ome_1 - 2 \int_0^{\infty} dk'
  \frac{\lam^2 v_{k'}^2}{\ome- k' \pm i \eps}. \EQN{2-9}
  \eeqa

With our form factor and small $\lam$,  Green's function
 has one pole $z_1$  in the lower half plane, i.e., 
$\eta^+(z_1)=0$ for 
\begin{equation}
 z_1 \equiv \omet_1 - i\gamma_1
 \EQN{SUM2'}
\end{equation}
The negative imaginary part  $-i\gamma_1$ describes decay for $t>0$.
The real part $\omet_1$ gives the shifted average energy of the excited state. For the other branch we have $\eta^-(z_1^*)=0$, with $z_1^*$ describing decay for $t<0$.

Note that in the representation (\ref{FreH}) the decay rate and shifted energy
of the excited state do not appear in the spectrum. One can incorporate $z_1$
(or $z_1^*$) into the spectrum by extracting the residue of \Ep{FreH} at the pole $\ome_k=z_1$ (or $z_1^*$). This gives the complex spectral decompositions \cite{PPT,Nakanishi,Sudarshan,Bohm}:
\begin{eqnarray}
  H  &=& |\phi_1\ket z_1 \bra\phit_1| +  \sum_k|\phi_k^+\ket \ome_k \bra\phit_k^+|  \nonumber\\
       &=& |\phit_1\ket z_1^* \bra\phi_1| +  \sum_k|\phi_k^-\ket \ome_k \bra\phit_k^-|
  \EQN{SUM2}
\end{eqnarray}
The state $|\phi_1\ket$ and its dual $\bra\phit_1|$ are complex eigenstates of $H$.
Their explicit forms are  \cite{PPT}
\begin{eqnarray}
 &&|\phi_1\ket
       = N_1^{1/2}
   \Big[|1\ket + \sum_k |k\ket {\lam V_k\over (z_1-\ome_k)^+}\Big],
 \EQN{phia}\\
 &&|\phit_1\ket
 = [N_1^*]^{1/2}
   \Big[|1\ket + \sum_k |k\ket {\lam V_k\over (z_1^*-\ome_k)^-}\Big].
 \nonumber
 \end{eqnarray}
The state $|\phi_k^+\ket$ has the same form as the state $|\phit_k^+\ket$ with the replacement
\begin{equation}
\frac{1}{\eta^+(\ome_k)} \Rightarrow \frac{1}{\eta^+(\ome_k)}  \frac{\ome_k - z_1}{(\ome_k-z_1)^+}
 \EQN{etad}
\end{equation}
and similarly, the state $|\phi_k^-\ket$ has the same form as the state $|\phit_k^-\ket$ with the complex conjugate replacement.

The states in \Ep{SUM2} form a bi-orthormal set,  with the relations
\begin{eqnarray}
  & &\bra\phit_1|\phi_1 \ket  = 1, \;\; \bra \phit_k^+| \phi_1\ket =0
  \nonumber \\
  & &\bra \phit_k^+| \phi_{k'}^+ \ket = \delta_{k,k'}
    \EQN{biort}
\end{eqnarray}
and their complex-conjugate relations.
%%%%%%%%%%%%%%%%%%%%
% Section 3 %
%%%%%%%%%%%%%%%%%%%%
\section{Two-atom system}
\label{sec:two}

\begin{figure}[htb] % Imported eps example.
\begin{center}
\includegraphics[width=2.5in]{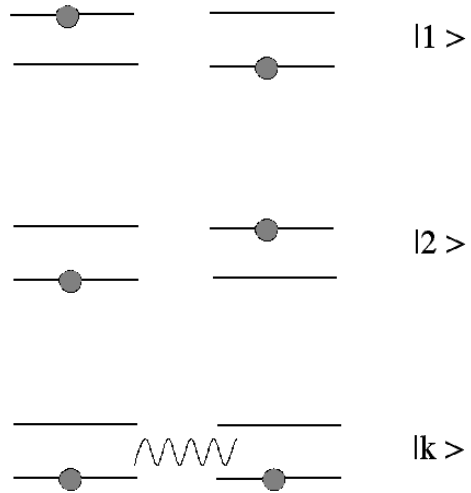} \caption{Two-atom system} \label{2fig}
\end{center}
\end{figure}

In this Section we discuss the complex spectral representation of a two-atom system  with Hamiltonian
\begin{eqnarray}
  H &=& \ome_1 |1\ket \bra 1| +  \ome_2 |2\ket \bra 2| +  \sum_k \ome_k |k\ket \bra k| \nonumber\\
  &+& \sum_k \lam_1 V_k \left(e^{i k x_1} |1\ket \bra k| + e^{-i k x_1} | k\ket \bra 1| \right) \nonumber\\
   &+& \sum_k \lam_2 V_k \left(e^{i k x_2} |2\ket \bra k| + e^{-i k x_2} | k \ket \bra 2| \right)
 \EQN{H2}
\end{eqnarray}
The state $|1\ket$ represents atom 1 in its excited state, while
atom 2 is in the ground state and no field is present. Conversely,
the state $|2\ket$ represents atom 2 in its excited state, while
atom 1 is in the ground state and no field is present. The state
$|k\ket$ represents a field mode $k$ with both atom 1 and
atom 2 are in their ground states  (see Figure \ref{2fig}). The atoms 1 and 2 are located at the positions  $x_1$ and $x_2$, respectively.
We use the potential in \Ep{Vomeka}.

We will first assume that
 the two atoms are at fixed positions, so that the distance between them
\beqa
 x_{21} = |x_2 - x_1|
 \EQN{x21def}
\eeqa
 is fixed. This can happen
if the atoms are  heavy. A system of two fixed atoms is analogous to a two-cavity waveguide (see Sec. \ref{wg}).
 
We will consider the case where the two atoms are identical,
\begin{eqnarray}
  && \lam_1 = \lam_2 = \lam \nonumber\\
  && \ome_1 = \ome_2
 \EQN{ident}
\end{eqnarray}
We introduce the symmetric and antisymmetric states
\begin{eqnarray}
  |s\ket = \frac{|1\ket + |2\ket}{\sqrt{2}}, \qquad  |a\ket = \frac{|1\ket - |2\ket}{\sqrt{2}}
   \EQN{pmstates}
\end{eqnarray}
which are eigenstates of the unperturbed Hamiltonian $H_0 = H(\lam=0)$ as
\begin{eqnarray}
  & &H_0 |j\ket = \ome_j |j\ket, \quad j=s,a \\
  && \ome_s = \ome_a = \ome_1.
     \EQN{H0pm}
\end{eqnarray}

We will use the notation 
\begin{eqnarray}
&&\sigma_j \equiv \left\{ \begin{array}{ll}
                  1, \,\, & \mbox{for $j=s$} \\
                  -1,    \,\, & \mbox{for $j=a$,} \end{array}\right.
\end{eqnarray}
With this notation we have $|j\ket = (|1\ket + \sigma_j |2\ket)/\sqrt{2}$.

As for the one-atom system, we can diagonalize the Hamiltonian as
 \begin{equation}
  H = \sum_k| \Ft_k^\pm\ket \ome_k \bra \Ft_k^\pm|
 \EQN{FreH2}
\end{equation}
where
 \begin{equation}
  | \Ft_k^\pm\ket = |k\ket + \beta^\pm_{sk} |s\ket + \beta^\pm_{ak} |a\ket + \sum_{k'} \beta^\pm_{k'k}
  |k'\ket
 \EQN{Fk}
\end{equation}
and
 \begin{equation}
 \beta^\pm_{jk}  =  \frac{1}{\sqrt{2}} \frac{\lam V_k}{\eta_j^\pm(\ome_k)} \left(e^{ikx_1} + \sigma_j  e^{ikx_2}\right)
 \EQN{bkp}
\end{equation}
 \begin{equation}
\beta^\pm_{k'k} =  \frac{1}{\sqrt{2}} \frac{\lam V_k}{\ome_k - \ome_{k'} \pm i\eps}
\sum_{j=s,a} \beta^\pm_{jk} \left(e^{-ik'x_1} + \sigma_j
e^{-ik'x_2}\right)
 \EQN{bkk}
\end{equation}
 \begin{eqnarray}
 & & \eta^\pm_j(\ome) \nonumber \\
 & &=  \ome - \ome_1 - 2\int_0^\infty dk' \frac{\lam^2 v_{k'}^2}{(\ome - k')^\pm }
 \left(1 + \sigma_j  \cos(k'x_{21})\right) \nonumber\\
 \EQN{nkp}
\end{eqnarray}
 Following a procedure similar to one found in Ref. \cite{PPT}, one can show that  the new diagonalized states $|\Ft_{k}^{\pm} \ket$ satisfy the orthogonality and completeness relations 
 \beqa
 & &\sum_k |\Ft_{k}^{\pm} \ket \bra \Ft_{k}^{\pm}| = |1\ket\bra 1| +
 |2\ket\bra 2| + \sum_{k} |k\ket\bra k| \EQN{eq2-1} \\
 & &\bra \Ft_{k}^{+} |\Ft_{k'}^{+}\ket = \bra \Ft_{k}^{-}
|\Ft_{k'}^{-}\ket  = \delta_{k, k'} \EQN{eq2-2}
 \eeqa

Green's function $[\eta^+_j(\ome)]^{-1}$ has poles in the lower half-plane, and conversely, $[\eta^-_j(\ome)]^{-1}$ has poles in the upper half-plane. From now on we discuss only the $+$ branch with poles on the lower half-plane. 

\begin{figure}[htb] % Imported eps example.
\begin{center}
\includegraphics[width=2.5in]{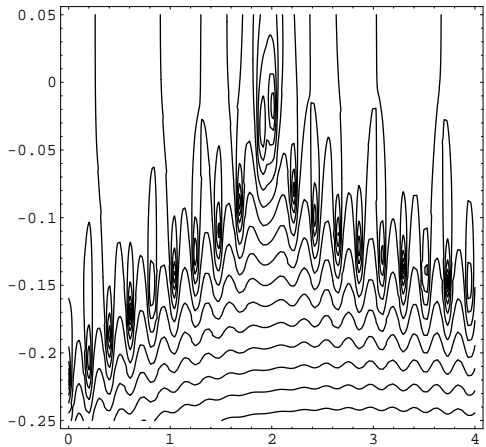}
\caption{Contour plot of  $\log (1/|\eta_s^+ (z)|)$. The $x$ and $y$
axes are ${\rm Re}(z)$ and ${\rm Im}(z)$, respectively. The contours  concentrate around the  poles $z_{s,n}$ of Green's function. Parameters are
$x_{21} = 29.025$, $\ome_1 = 2$, $\lam=0.05$, $\ome_M = 5$.} 
\label{polecontour} \end{center}
\end{figure}

A new feature with respect to the one-atom system is that due to the cosine term in \Ep{nkp}, there are many poles of Green's function, as shown in Fig. \ref{polecontour}.  We label the poles as
 \beqa
 z_{j,n} = \omet_{j,n} - i\gamma_{j,n} 
 \EQN{zjn0}
 \eeqa
where $n$ is an integer. 

The poles $z_{j,n}$ are solutions of the equation
 \beqa
 \eta^+_j(z_{j,n}) = 0 
 \EQN{eta0}
 \eeqa
In the following we discuss this equation and its solutions.

From  Eqs. (\ref{complexd})  and (\ref{nkp}) we obtain
 \begin{eqnarray}
 z_{j,n}
 &=&  \ome_1 +  2\int_0^\infty dk' \frac{\lam^2 v_{k'}^2}{z_{j,n} - k' }
 \left(1 + \sigma_j  \cos(k'x_{21})\right) \nonumber\\
 &-& 4\pi i \lam^2 [v_{z_{j,n}}]^2  \left(1 + \sigma_j  \cos(z_{j,n} x_{21})\right)
 \EQN{expg}
\end{eqnarray}
Note that cosine in the last term includes the factor $\exp(\gamma_{j,n} x_{21})$, which grows exponentially with the distance $x_{21}$ between the atoms. 
Assuming  weak coupling and taking only the  pole contribution in the $k'$ integral we obtain the set of equations
 \beqa
  \omet_{j,n} \approx \ome_1 + 2\pi  [\lam v(\omet_{j,n})]^2 \sigma _j e^{\gamma_{j,n} x_{21}} \sin(\omet_{j,n} x_{21})
  \EQN{many1}
  \eeqa
 \beqa
  \gam_{j,n} \approx  2\pi  [\lam v(\omet_{j,n})]^2 \left[1 + \sigma _j e^{\gamma_{j,n} x_{21}} \cos(\omet_{j,n} x_{21})\right]
  \EQN{many2}
  \eeqa
In Fig. \ref{polecontour}, the pole of $[\eta^+_s(\ome)]^{-1}$ with real part closest to the unperturbed frequency $\ome_1$ is also closest to the real axis. We call this pole $z_{s,0} = z_s$. Similarly, for $[\eta^+_a(\ome)]^{-1}$ we denote   the pole closest to $\ome_1$ as $z_{a,0} = z_a$.  Both these poles are obtained by a perturbation expansion around  $\lam=0$. We have 
 \beqa
 z_{j,0} = z_j \to \ome_1 \quad {\rm as\,\,} \lam \to 0.
  \EQN{zjpert}
  \eeqa
 If $x_{21}$ is not too large ($x_{21} \sim \gam_1^{-1}$ or smaller),  one can show that the poles $z_{j,n}$  are given by 
\begin{eqnarray}
&& z_{j,n} \approx \left\{ \begin{array}{ll}
                  z_j + 2n\pi  /x_{21}  + \del z_{j,n}, \,\, & \mbox{for $\sigma_j n > 0$} \\
                  z_j + (2 n+\sigma_j)\pi /x_{21}  + \del z_{j,n},    \,\, & \mbox{for $\sigma_j n < 0$.}       
                    \end{array}\right.\nonumber\\
                    \EQN{many3}
\end{eqnarray}
where $\del z_{j,n}$ is an $O(\lam^2$) correction.  The approximate value ${\rm Re}( z_{j,n} - z_j)$ predicted by this equation agrees with  Fig. \ref{polecontour} (for $j=s$) and a similar figure for $j=a$, which we omit.

We write the poles $z_j$ as
\begin{eqnarray}
   z_j = \omet_{j} - i \gamma_{j}.
     \EQN{ojgj}
\end{eqnarray}
The poles $z_j$, having the smallest decay rates $\gamma_j$, will give a dominant contribution to the time evolution after a few bounces of the field between the atoms. In this way, the complex collective states defined in \Ep{Hpm} emerge.

As in the one-atom system we can obtain complex   eigenstates $|\phi_j \ket$ of the 
 total Hamiltonian, such that 
\begin{eqnarray}
  H |\phi_j\ket = z_j |\phi_j\ket
       \EQN{Hpm}
\end{eqnarray}

Their explicit forms  are given by 
\begin{eqnarray}
&& |\phi_j\ket \nonumber \\
&&=N_j^{1/2} \left[ |j\ket +\sum_k \frac{2^{-1/2} \lam V_k}
 {(z_j - \ome_k)^+} \left(e^{-ikx_1} + \sigma_j e^{-ikx_2}\right)|k\ket \right] \nonumber\\
&&\EQN{fpm}
\end{eqnarray}
 where
 \beqa
 N_j = \left[ 1+ \sum_{k} \frac{\lam^2 V_{k}^2}{ [(z_j -\ome_k)^+]^2}
 (1+ \sigma_j \cos k x_{21} ) \right]^{-1}. \EQN{eq2-3}
 \eeqa
For these states we have $|\phi_j\ket \rightarrow |j\ket$ as $\lam
 \rightarrow 0$.

The dual  states satisfying $\bra \phit_j|H = z_j\bra \phit_j|$ are given by
 \beqa
 & &\bra \phit_j | = N_j^{1/2} \left[ \bra j| + \sum_k \frac{
 2^{-1/2} \lam V_{k} }{(z_j - \ome_k)^+} \left(e^{ikx_1} + \sigma_j
 e^{ikx_2}\right)\bra k| \right] \nonumber\\ 
\EQN{eq2-4}
 \eeqa
 Like in the one-atom case, we have the complex spectral
 representation
 \beqa
  H = \sum_{j=s,a} |\phi_j\ket z_j \bra \phit_j| + \sum_{k}
  |F_k^+\ket \ome_k \bra \Ft_{k}^+| \EQN{eq2-6}
  \eeqa
 where $|F_k^{+} \ket $ has the same form as the state
 $|\Ft_k^{+}\ket$ with the replacement
 \beqa
 \frac{1}{\eta^{+}_j(\ome_k)} \Rightarrow \frac{1}{\eta^{+}_j(\ome_k)}
 \frac{\ome_k - z_j}{(\ome_k -z_j)^{+}} \EQN{eq2-7}
 \eeqa
We have as well the complex-conjugate representation, taking the complex-conjugates of Eqs. (\ref{eq2-6}), (\ref{eq2-7}).

%%%%%%%%%%%%%%%%%%%%
% Section 3 %
%%%%%%%%%%%%%%%%%%%%
\section{Emergence of the complex collective  states}
\label{sec:emergence}

Time evolution of the two atom system can be solved by using
\Ep{FreH2} or \Ep{eq2-6}. As an example we assume the atoms are initially in the symmetric
 state $|s\ket$ and the initial field is zero (similar calculations can be done if the initial state is $|a\ket$).   We will calculate the survival probability of state
  $|1\ket$,
\beqa
 P_1 (t) =  |\bra 1| e^{-i H t}|s\ket |^2
\EQN{P1def}
\eeqa
Before we go into details, we can guess the 
 behavior the system will show. Since the initial state is symmetric, the following discussion also applies with   atoms 1 and 2 exchanged. Say atom 1 is to the left of atom 2. At the beginning, atom 1  decays
 and emits a field. Half of this field will be radiated away to the left, while the other half will reach and excite atom 2. Atom 2 will then decay and emit its own field, part of which will be radiated away to the right, the rest going to the left, back towards atom 1. Continuing this process, we see that  energy will bounce back and forth between the two atoms. As time passes, this energy will decrease due to the outgoing radiation. Eventually both atoms will decay to the ground state.  Noting that the time it takes for the field of one atom to reach the other atom is  $t=x_{21}$  (with $c=1$) we conclude that, as it decreases,  the survival probability should oscillate with period $x_{21}$.

\begin{figure}
\centering
\includegraphics[width=2.4in,angle=270]{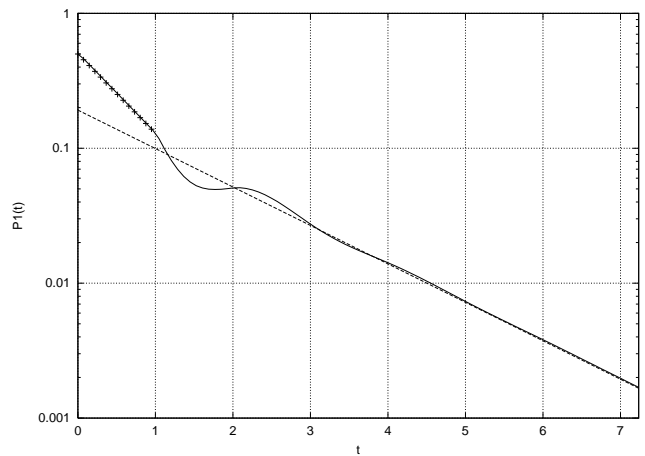}
 \caption{Log plot of the survival probability $P_1(t)$ (solid line) and  complex collective  state component $P_{1,zs}(t)$ (dashed line).  The crosses indicate the decay rate $2\gam_{s1}$, $\gam_{s1} = 0.0233$, between $t=0$ and $t=1$ (see Sec. \ref{sec:b}). Time is in units of $x_{21}$. The distance between atoms is  $x_{21} = 29.025$. Other parameters are  $\ome_1=2$, $\lam=0.05$, $\ome_M=5$, and  $L=500$.}
  \label{f:p1t} 
 \end{figure}

This behavior is shown in Fig. \ref{f:p1t}. This was obtained through a numerical solution of Schr\"odinger's equation. The field was
discretized into $2501$ modes. The eigenvalues and eigenfunctions of the Hamiltonian matrix were obtained using  tri-diagonalization and the ``QL'' method \cite{EIS}.  This allowed us to calculate explicitly the operator $\exp(-iHt)$. For this and the subsequent numerical plots we used the following parameters: $\ome_1=2$, $\lam=0.05$, $\ome_M=5$. Other parameters are indicated in each figure.

In order to calculate the survival probability we  start with \Ep{FreH2} to obtain 
  \beqa
  P_1 (t) &=& |\bra 1| e^{-i H t}\sum_{k}|\Ft_{k}^+\ket \bra \Ft_{k}^+|s\ket
  |^2 \nonumber \\
 &=& \left| \sum_{k} e^{-i \ome_k t} \bra 1|\Ft_{k}^+\ket\bra \Ft_{k}^+|s\ket
  \right|^2 \nonumber \\
 &=& \frac{1}{2}\left| \sum_{k}e^{-i \ome_k t} \frac{\lam^2 V_k^2}{|\eta_s^+
 (\ome_k)|^2} (1+\cos kx_{21}) \right|^2 \EQN{eq3-1}
  \eeqa
where used the fact that odd functions of $k$ vanish under the summation. For later use we define the amplitude in \Ep{eq3-1} as
  \beqa
  I (t) \equiv \sum_{k}e^{-i \ome_k t} \frac{\lam^2 V_k^2}{|\eta_s^+
 (\ome_k)|^2} (1+\cos kx_{21}) 
  \EQN{Itdef}
  \eeqa

The dominant contribution to $P_1(t)$ will come from the poles of Green's function, shown
in Fig. \ref{polecontour}. The different pole contributions should add up to give the bounces seen in Fig. \ref{f:p1t}.  But rather than computing  all the pole contributions,  we will follow an easier method in Sec. \ref{sec:b}. 

Here we will focus on the pole $z_s$. As mentioned before, this will give the dominant contribution after some bounces, since it gives the slowest decay rate. It is this pole contribution that is extracted in the representation (\ref{eq2-6}). Using this representation, and noting that $\bra \phit_a|s\ket = 0$ we have
\beqa
P_1(t)=  \left|\bra 1|\phi_s\ket e^{-i z_s t} \bra \phit_s|s\ket + \sum_k  \bra 1|F_k^+\ket e^{-i \ome_k t} \bra \Ft_k^+|s\ket \right|^2\nonumber\\
 \EQN{P1approx0}
\eeqa
 The second term contains contributions from the poles other than $z_s$ of Green's function $[\eta^+_s(\ome)]^{-1}$ as well as contributions coming from the branch cut of this function. Neglecting all these contributions we obtain
\beqa
P_1(t) \approx P_{1, zs} (t) \equiv \left|\bra 1|\phi_s\ket e^{-i z_s t} \bra \phit_s|s\ket \right|^2
 \EQN{P1approx}
\eeqa
This is represented by the dashed line in Fig. \ref{f:p1t}. After a few  bounces the initial state $|s\ket$ reaches the collective  state $|\phi_s\ket$.

\begin{figure}[htb] % Imported eps example.
\begin{center}
\includegraphics[width=2.4in,angle=270]{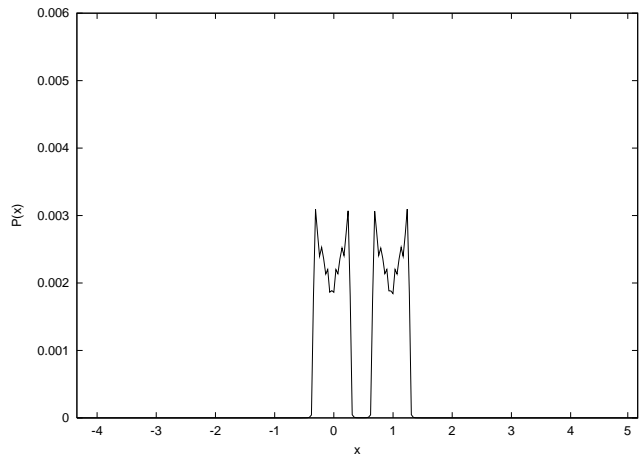} \caption{
Field intensity $P(x,t)$ for $t=0.32\, x_{21}$.  The atoms are located at $x_1=0$ and $x_2=1$. Space coordinate $x$ is in units of $x_{21}$ and $P(x,t)$ is dimensionless. The parameters are the same as in Fig. \ref{f:p1t}. } \label{plot1}
\end{center}
\end{figure}

\begin{figure}[htb] % Imported eps example.

\begin{center}
\includegraphics[width=2.4in,angle=270]{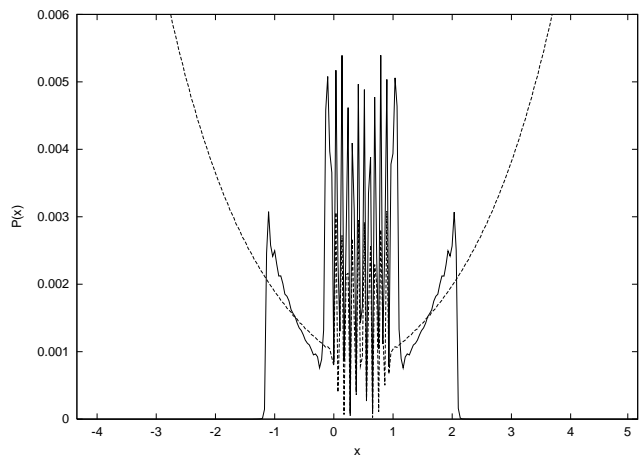} \caption{
Field intensity $P(x,t)$ for $t = 1.12\, x_{21}$ (solid line) and the complex collective  state  component $P_{zs} (x,t)$ (dashed line). Space coordinate $x$ is in units of $x_{21}$ and $P(x,t)$ is dimensionless. Parameters are the same as in Fig. \ref{f:p1t}. } \label{plot2}
\end{center}
\end{figure}

\begin{figure}[htb] % Imported eps example.
\begin{center}
\includegraphics[width=2.4in,angle=270]{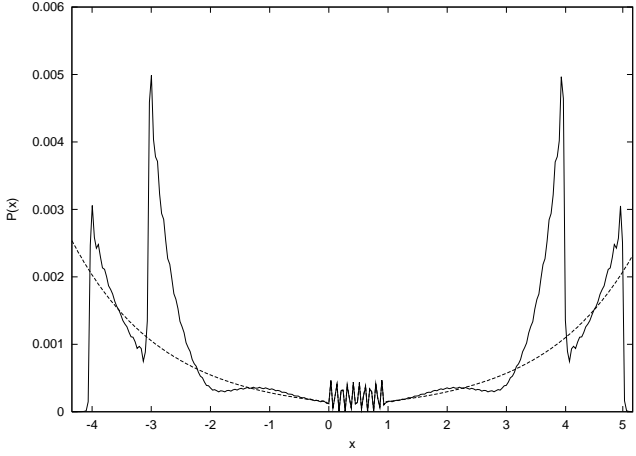}\caption{
Field intensity $P(x,t)$ for $t = 4.02\, x_{21}$ (solid line) and the complex collective  state component $P_{zs} (x,t)$ (dashed line).  The outer smaller peaks of $P(x,t)$ come from the initial one-atom emission. The
inner, larger peaks come from the emission after the first exchange of energy between the atoms.  $P(x,t)$ asymptotically approaches $P_{zs} (x,t)$ as $x$ approaches the atoms. 
They coincide in the region between the atoms (between $x=0$ and $x=1$).  Parameters are the same as in Fig. \ref{f:p1t}} \label{plot3}
\end{center}
\end{figure}

We turn  to the time evolution of the field. Defining the state
  \beqa
   |\psi(x) \ket = \sum_k \frac{1}{(2 \ome_k L)^{1/2}} e^{-i k x}
   |k\ket , \EQN{eq3-2}
   \eeqa
 the intensity of the field in space-time can be written as
  \beqa
  P(x,t) = |\bra \psi(x) | e^{-iH t} | s\ket|^2
  \EQN{eq3-3}
 \eeqa
Again we calculated this using the numerical solution  of Schr\"odinger's equation. 
 The intensity of the field is plotted in Figs. \ref{plot1}-\ref{plot3}
  for different times. At the beginning, both atoms  emit their fields spontaneously.
  Each field has an exponentially growing envelope (plus corrections due to the initial dressing processes \cite{POP2001}),  which stops at the light cone $|x-x_i| = t$ (Fig. \ref{plot1}).
  
 After each emitted field reaches the  neighbor atom, absorption
 and re-emission occur.  The two atoms exchange energy and the field $P(x,t)$ around the atoms starts to approach the  field intensity due to the collective state given by  
  \beqa
 P_{zs} (x,t) = | \bra \psi(x)|\phi_s\ket\exp(-iz_st)\bra\phi_s|s\ket|^2 
 \EQN{eq3-3'}
 \eeqa
(see Fig. \ref{plot2}). The collective state  decays exponentially (Fig. \ref{plot3}).


In summary, the atoms emit a field growing exponentially with the distance from them, within their light-cones. After the field emitted from each atom reaches the other one, the collective state with complex energy $z_j$  emerges. 

As we discuss now, the exponential field has a strong influence on $z_j$.  The field amplitude associated with the collective states is given by   $\bra k|\phi_j\ket$. This amplitude in turn determines $z_j$ through its interaction with the atoms. We have
 \begin{eqnarray}
 z_j 
=  \ome_1 +  \frac{1}{\sqrt{2 N_j}} \sum_k \lam V_k \left[e^{i kx_1}+ \sigma_j e^{ikx_2}\right] \bra k|\phi_j\ket
 \EQN{expg2}
\end{eqnarray}
where we used  Eqs. (\ref{Hpm}), (\ref{fpm}), and (\ref{nkp}), with $\ome=z_j$.
Since  $\bra k|\phi_j\ket$ are functions of $z_j$, this is a self-consistent relation.  The exponential component of the field is seen in the approximate equations (\ref{many1}) and  (\ref{many2}) for $z_{j,0} =z_j$, which include the factor $\exp(\gamma_j x_{21})$. 
Due to the exponential nature of this factor, the pole $z_j$ may deviate substantially from the one-atom pole $z_1$.

\begin{figure}
\begin{center}
\includegraphics[width=3.25in]{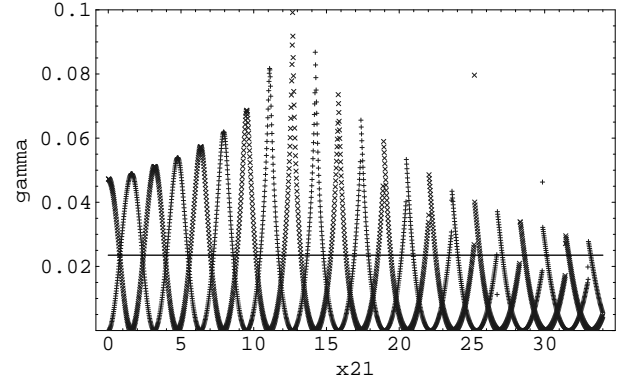} \caption{
The decay rates $\gam_s$ ($\times$) and $\gam_a$ ($+$) oscillating as a function of $x_{21}$. The solid line is the one-atom decay rate
$\gam_1 = 0.0235$. $\gam_s$ vanishes for distances close to $(2n-1)\pi/\omet_{1}$,   and $\gam_a$ for distances close to $2n\pi/\omet_1$ with $n$ integer. An example is  $x_{21}  = 12.7 \approx 8 \pi/\omet_1$, where $\gam_a$ vanishes, while $\gam_s$ is large. For large $x_{21}$ the decay rates decrease. } \label{gams}
\end{center}
\end{figure}

In spite of the exponential factor, for increasing $x_{21}$  the  equations (\ref{many1}) and  (\ref{many1}) can still have solutions if $\gamma_j$ decreases as
 \beqa
  \gamma_j \sim x_{21} ^{-1}
  \EQN{many4}
  \eeqa
for large $ x_{21}$. The decrease of $\gamma_j$ with increasing $x_{21}$ is seen in Fig. \ref{gams}.

%%%%%%%%%%%%%%%%%%%%%%%%
\section{Bounces}
\label{sec:b}
%%%%%%%%%%%%%%%%%%%%%%%%%%
In this section we describe the energy bounces between the atoms, seen in the survival probability of each atom.

As shown in Fig.  \ref{f:p1t}, the decay rate of $P_1(t)$ changes abruptly at $t=x_{21}$. For $t > x_{21}$
 the decay rate quickly approaches the collective decay rate
 $\gamma_s$. The wiggling of the decay rate shows the absorption and re-emission
 of the fields, or in other words the  energy bounces. For $0 < t<x_{21}$  the decay rate should be  close to the
 one-particle decay rate.  

To analyze the energy bounces and the decay for $0 < t<x_{21}$ we first note that
  \beqa
  \eta_s^+(k) -\eta_s^- (k) = 4\pi i \lam^2 v_{k}^2 (1+\cos
  kx_{21}), \EQN{eq3-1-1}
  \eeqa
Hence in \Ep{Itdef} we can write
 \beqa
  I(t)  &=& 2 \int_0^{\infty} dk
 e^{-i \ome_k t} \frac{\lam^2 v_k^2}{|\eta_s^+
 (k)|^2} (1+\cos kx_{21}) \nonumber \\
&=& \frac{1}{2\pi i} \int_0^{\infty} dk
 \left( \frac{1}{\eta_s^-(k)} - \frac{1}{\eta_s^+(k)}\right) e^{-i
 k t} \EQN{eq3-1-2}
 \eeqa

Since $e^{-i k t}$ vanishes in the lower  infinite semi-circle of
  complex $k$ plane for $t>0$, we can take the pole contributions  extending the $k$ integration from $-\infty$ to $\infty$ and closing the contour with this semi-circle. Only 
 $[\eta_s^+(k)]^{-1}$ has poles in the lower half-plane.  We write $\eta_s^+ (k)$ as
 \beqa
 \eta_s^+ (k) &=& k -\ome_1 - 2 \int_0^{\infty} dk'
 \frac{\lam^2 v_{k'}^2 (1+\cos k'x_{21})}{k-k'+i \eps} \nonumber \\
 &=& \eta_{s1}^+(k)  -\Del(k)  \EQN{eq3-1-3}
 \eeqa
 where $\eta_{s1}^+(k)$ is defined by
 \beqa
 & &\eta_{s1}^+(k)  \nonumber \\
 & &= k - \ome_1 - 2\int_0^{\infty} dk' \frac{\lam^2
 v_{k'}^2}{k-k'+i\eps}(1+\frac{1}{2}e^{-i k' x_{21}}) \nonumber \\
 & &- \int_0^{\infty} dk' \frac{\lam^2v_{k'}^2 e^{i k'
 x_{21}}}{k-k'-i\eps}. \EQN{eq3-1-4}
 \eeqa
and 
\beqa
 \Del(k) = - 2\pi i \lam^2 v_k^2 e^{ikx_{21}}
\EQN{Delk}
\eeqa
 Unlike $[\eta_{s}^+(k)]^{-1}$,  $[\eta_{s1}^+(k)]^{-1}$ has only one pole
 in the lower half plane. Let this pole  be
 \beqa
 z_{s1} =  \omet_{s1} - i \gam_{s1} 
 \EQN{zs1def}
 \eeqa
This is essentially the pole of one-atom Green's function, modified by the overlap of the atomic clouds at the distance $x_{21}$. 
For $x_{21}\gg \ome_1^{-1}$ we have
 \beqa
 z_{s1} \approx z_1.
 \EQN{zs1z1}
 \eeqa
  We expand $1/\eta_{s}^+(k)$ as 
  \beqa
  & &\frac{1}{\eta_{s}^+(k)} = \frac{1}{\eta_{s1}^+(k) - \Del(k)} \nonumber \\
  & &= \sum_{n=0}^{\infty} \frac{ \Del(k)^n}{ (\eta_{s1}^+ (k) )^{n+1}}.\EQN{eq3-1-5}
  \eeqa
 The expansion is possible since
  \beqa
   & &\eta_{s1}^+(k) = k- \ome_1 - 2\int_0^{\infty} dk'\,{\cal P} \left(
   \frac{\lam^2 v_{k'}^2 (1+\cos k'x_{21})}{k-k'} \right) \nonumber \\
   & &+ 2 \pi
   \lam^2 v_{k}^2 \sin k x_{21} + 2 \pi i \lam^2 v_{k}^2,
   \eeqa
   \beqa
  |\Del(k)| = |{\rm Im} (\eta_{s1}^+(k))|   \le | (\eta_{s1}^+(k)) |.
   \EQN{3-1-6}
 \eeqa
 Using \Ep{eq3-1-5}, \Ep{eq3-1-2} is written as
  \beqa
  & &I(t) = \frac{1}{2\pi i} \int_0^{\infty} dk
 \frac{e^{-ikt}}{\eta_s^-(k)} \nonumber \\
 & &- \frac{1}{2\pi i} \sum_{n=0}^{\infty} \int_0^{\infty} dk\,
 \frac{(-2\pi i \lam^2 v_{k}^2)^n e^{-i k
 (t-nx_{21})}}{(\eta_{s1}^+(k))^{n+1}}. \EQN{3-1-7}
  \eeqa
In \Ep{3-1-7}, the pole contributions come from
 $1/(\eta_{s1}^+(k))^{n+1}$. For $n=0$, $e^{-ikt}/\eta_{s1}^+(k)$ has a
 simple pole in the lower half plane at $k=z_{s1}$. Its effect appears for $t>0$, when we can  close the
 integration contour in the lower half plane. For $n=1$, $e^{-ik(t-x_{21})}/(\eta_{s1}^+(k))^2$
 has a double pole. Its effect appears  for $t>x_{21}$. In general, for each $x_{21}$ time step there appears a new pole effect which is smaller by $\lam^2$ order than the previous pole
 effect. In this way we can explain the wiggling decay rate (Figure
 \ref{f:p1t}).

As we discuss now, this description of the bounces is connected to emergence of the collective state. Approximating (for $\lam\ll 1$)
\beqa
  \eta^+_{s1}(k)  \approx k - z_{s1}
\EQN{Delk3}
\eeqa
the pole contributions in \Ep{3-1-7} are given by
\beqa
 I_0(t) &\approx&  \frac{-1}{2\pi i}  \int_{-\infty}^{\infty} dk\,
 \frac{1 }
 {k-z_{s1} - \Del(k)} e^{-i k t} \nonumber\\
   &=& \frac{-1}{2\pi i} \sum_{n=0}^{\infty} \int_{-\infty}^{\infty} dk\,
 \frac{\Del(k)^n }
 {(k-z_{s1})^{n+1}}e^{-i k t}
\EQN{Delk3'}
\eeqa
Taking the residues at the pole $k=z_{s1}$ we obtain an expression of the form
\beqa
 I_0(t) =  \sum_{n=0}^{\infty} \theta(t- n x_{21}) f_n(t) 
  \EQN{bces}
\eeqa
where 
\beqa
 f_n(t) = - \left[\frac{1}{n!} \frac{\pard^n}{\pard k^n} \Del(k)^n e^{-ikt} \right]_{k=z_{s1}}.
  \EQN{fndef}
\eeqa
We have $f_n \sim \lam^{2n}$. Note that the sum stops at $n$ such that $t< n x_{21}$. Since for weak coupling the terms $f_n(t)$ become smaller  as $n$ increases,  after a few bounces we have
\beqa
  I_0(t)  \approx  {\It}_0(t)
  \EQN{bces2}
\eeqa
where
\beqa
 {\It}_0(t) =  \sum_{n=0}^{\infty}  f_n(t) 
  \EQN{bces3}
\eeqa
As shown in Appendix \ref{app:zszs1} we have
\beqa
  {\It}_0(t) =  N_s e^{-iz_s t}
  \EQN{bces6}
\eeqa
for all $t>0$, where 
\beqa
  N_s = \frac{1}{1- \pard \Del(k) /\pard k}\Big|_{k=z_{s}}
  \EQN{Ns'}
\eeqa
for weak coupling. \Ep{bces6} shows  that the sum of all bounces gives the contribution  from the collective state $|\phi_s\ket$ with eigenvalue $z_s$. 

\Ep{bces6} is consistent with  $z_s$ giving the slowest exponential decay. To see this
we use \Ep{eq3-1-3} to write the equation for $z_{s,n}$ as
\beqa
 \eta^+_{s1}(z_{s,n}) - \Del(z_{s,n}) = 0
\EQN{Delk2}
\eeqa
or 
\beqa
 z_{s,n} \approx z_{s1} + \Del(z_{s,n})
\EQN{Delk4}
\eeqa
[for $x_{21} \gg \ome_1^{-1}$ we have $z_{s1} \approx z_1$ and we recover Eqs.  (\ref{many1}) and (\ref{many2}) for $j=s$]. 

The function $k-z_{s1} - \Del(k)$ in \Ep{Delk3'} has zeroes at  $k = z_{s,n}$.  For $t\to\infty$ only the residue at the pole $k=z_{s,0}=z_s$ remains and we get
\beqa
  \lim_{t\to\infty}  N_s^{-1}e^{i z_s t} I_0(t) =  \lim_{t\to\infty}  N_s^{-1} e^{i z_s t} \It_0(t)  = 1
  \EQN{bces4}
\eeqa
which is consistent with \Ep{bces6}.

\begin{figure}
\begin{center}
\includegraphics[width=3.25in]{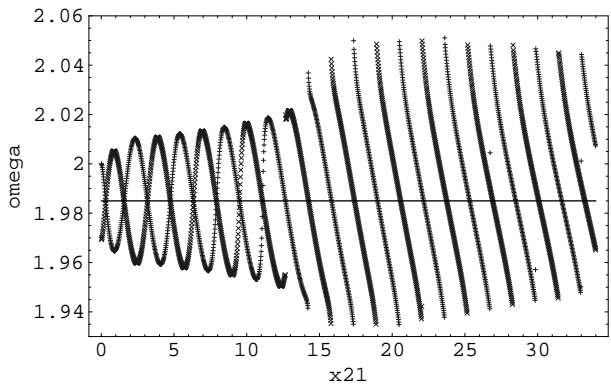}
 \caption{The energies $\omet_s$ ($\times$) and $\omet_a$ ($+$) as a function of $x_{21}$.  They
 oscillate with $x_{21}$. The solid line is the one-atom energy $\omet_1 = 1.985$.} \label{omes}
 \end{center}
 \end{figure}

%%%%%%%%%%%%%%%%%%%%
% Section 3 %
%%%%%%%%%%%%%%%%%%%%
\section{Decay rate and energy vs. distance}
\label{sec:x12}

\begin{figure}
\begin{center}
\includegraphics[width=2.4in,angle=270]{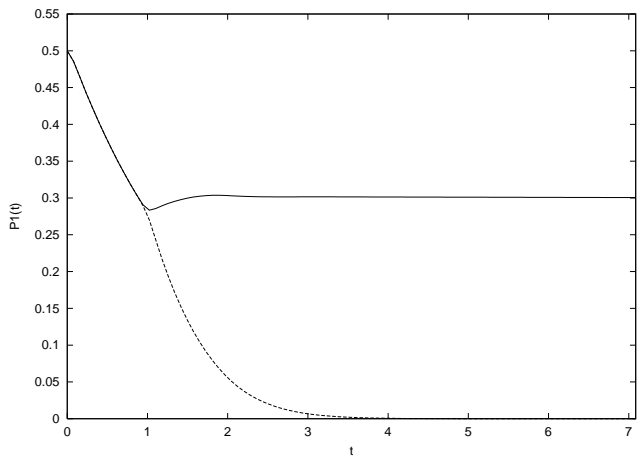}
 \caption{Survival probability $P_1(t)$ of atom 1 for symmetric (dashed) and antisymmetric (solid) initial conditions and $x_{21} = 12.7$. The symmetric state gives rise to a super-radiant, stationary collective state. The antisymmetric state gives rise to a sub-radiant (stationary) collective state. Before $t=x_{21}$, $P_1(t)$ has the one-atom decay rate. Time $t$ is in units of $x_{21}$. $P_1(t)$ is dimensionless.} \label{pmsurv}
 \end{center}
 \end{figure}
 
\begin{figure}
\begin{center}
\includegraphics[width=2.4in,angle=270]{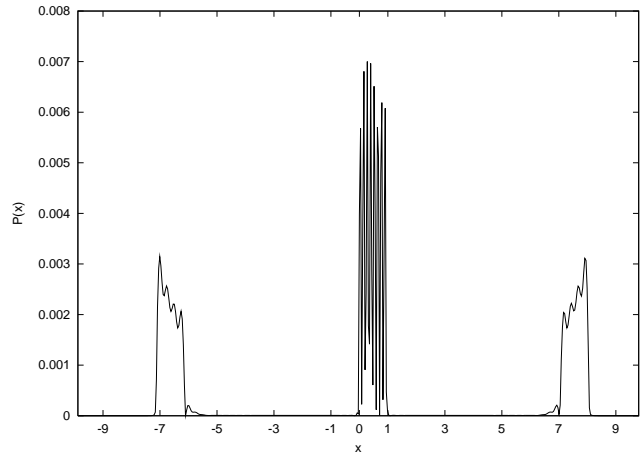}
 \caption{Field intensity $P(x,t)$ for the atoms in a stationary collective state. The field between the two atoms, located at $x=0$ and $x=1$ remains trapped. $x$ is in units of $x_{21} =12.7$ and $P(x,t)$ is dimensionless.  The initial condition is $|a\ket$. Time is $t=7.09\, x_{21}$.  The wave packets on each side represent the field emitted before the atoms formed the collective state. After this, there is no emission (we have sub-radiance).} \label{Pxm}
 \end{center}
 \end{figure}

 \begin{figure}
\begin{center}
\includegraphics[width=2.4in,angle=270]{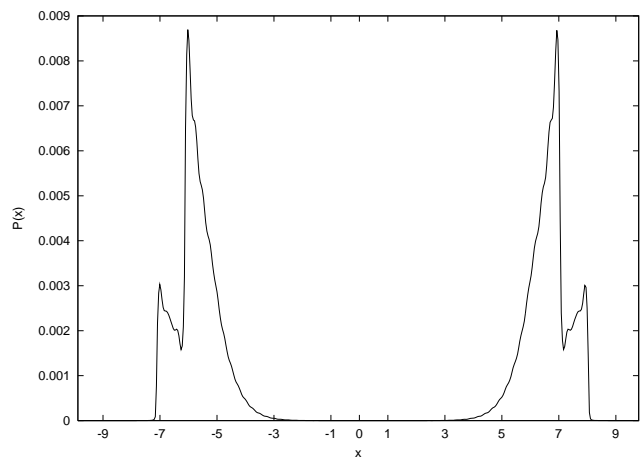}
 \caption{Field intensity $P(x,t)$ for  initial condition $|s\ket$ at time  $t=7.09\, x_{21}$. 
$x$ is in units of $x_{21}=12.7$ and $P(x,t)$ is dimensionless. The collective  state has decayed. The smaller peaks on the far  sides are the field emitted individually  by each atom before attaining the collective state. The larger peaks  correspond to the two-atom collective emission (super-radiance).} \label{Pxp}
 \end{center}
 \end{figure}

 \begin{figure}
\begin{center}
\includegraphics[width=3.25in]{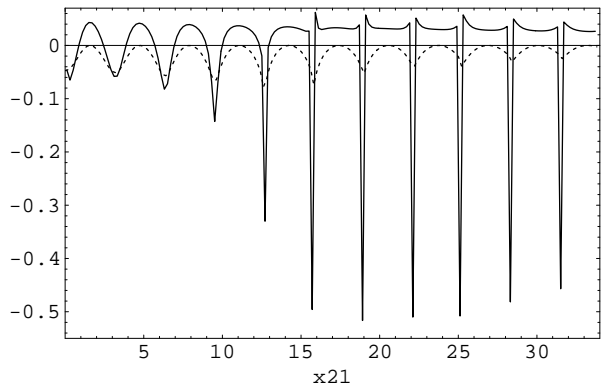}
\caption{ The graph of  ${\cal F}_s = -d\omet_s/dx_{21}$  (black line) and the decay rate $- \gam_s$  (dotted line). We see that when $-\gam_s$ is a local minimum ${\cal F}_s$ has also a local minimum value.} \label{Fng}
\end{center}
\end{figure}

In this section we investigate the behavior of the complex eigenvalues of the Hamiltonian $z_j$  for different values of $x_{21}$.

The equation $\eta^+_j(z) =0$ can be solved
numerically by iterations of $z= z - \eta^+_j(z)$. The imaginary and real  parts of $z$ thus obtained are shown in  Figs. \ref{gams} and \ref{omes} (we used the same parameters as in the previous figures). The numerical iteration was started around $z=\ome_1$ so, with the exception of two isolated points seen in Fig. \ref{gams}, the solutions obtained are the collective eigenvalues $z=z_j$. Gaps in the graphs are points missed by the numerical solution.

As we  see,  $\gamma_j$ and $\omet_j$ oscillate with $x_{21}$. The oscillation period
 is  approximately $2\pi/\omet_1$ where $\omet_1$ is the one-atom renormalized frequency (see Appendix \ref{app:gamma}). 
 
Due to the oscillations, the collective decay rate can become smaller or larger than the one-atom decay rate (solid line in Fig. \ref{gams}).  We have sub-radiance and super-radiance, respectively. In particular, it is noticeable  that there are distances at which the decay rates $\gamma_j$ vanish  (see Appendix \ref{app:gamma}). This means that for these distances there is no outgoing radiation. The outgoing emitted fields of the atoms cancel by destructive interference and a standing field is trapped  between
 two atoms, storing energy. Note that both symmetric and antisymmetric initial conditions can give rise to either sub-radiant or super-radiant states, since both $\gam_s$ and $\gam_a$ oscillate with $x_{21}$.

The oscillations of the decay rate and the energy shown in Figs. \ref{gams} and \ref{omes} are a unique feature of  one-dimensional systems. For two or three dimensions, these quantities can change significantly only for short distances between atoms (see Appendix \ref{app:23}).

As an example of sub-radiance and super-radiance we show numerical simulations with the same parameters used in the previous examples, except we choose $L=250$ (to have higher space resolution) and $x_{21}=12.7$. For this value of $x_{21}$, the decay rate of the antisymmetric state vanishes while the decay rate of the  symmetric state is maximum,  (see Fig. \ref{gams}). In Fig. \ref{pmsurv} we show the survival probability of atom 1 for the antisymmetric  and symmetric initial conditions, showing the appearance of stationary sub-radiant collective state and a super-radiant collective  state. In Figs. \ref{Pxm} and  \ref{Pxp} we show the corresponding fields.

We turn to the force between the atoms. Here we will only give a  heuristic discussion. A more detailed analysis requires including the Casimir-Polder or van der Waals forces between the atoms, as well as the inertia of the atoms, which we are not considering in this paper. 
 
 Since the atoms are unstable, the force between them should be time-dependent \cite{Passante}. 
 We expect  the force will decay exponentially during the time scales where the collective-state components dominate. For the dependence on $x_{21}$  of the force, the quantity
 \beqa
 {\cal F}_j = -d\omet_j /dx_{21}
 \EQN{forcedef}
 \eeqa
can give an indication because $\omet_j$ is the average energy of the collective state.  

As we can see in 
 Figure \ref{Fng},  ${\cal F}_s$ oscillates with  $x_{21}$ (${\cal F}_a$ has a similar behavior). ${\cal F}_s > 0$ corresponds to a repulsive force, and ${\cal F}_s <0$ to an attractive force. The attractive force between two atoms becomes locally maximum
 when the collective decay rate is locally maximum.  The atoms tend to attract each other when they emit the field outwards, and tend to repel when the field remains trapped between them.
 
 Also we see in Fig. \ref{Fng} that there are points $x_{21}^{0}$   for which ${\cal F}_s$ vanishes. If $d{\cal F}_s/dx_{21} <0$ at these points, then any small displacement $\Del x_{21}$ around $x_{21}^{0}$ creates a force in the opposite direction. Thus in this case $x_{21}^{0}$are stable points [if $d{\cal F}_s/dx_{21} >0$ the points are unstable].  The existence of stable points  suggests the possibility of having a one-dimensional ``molecule.''  This molecule  would have a  lifetime of the order of $\gamma_s^{-1}$.

%%%%%%%%%%%%%%%%%%%%
% Section 6 %
%%%%%%%%%%%%%%%%%%%%

\section{Two-cavity wave guides}
\label{wg}

\begin{figure}[htb] % Imported eps example.
\begin{center}
\includegraphics[width=2.5in]{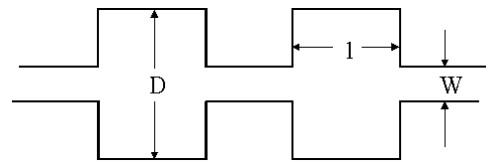}
 \caption{A two-cavity waveguide} \label{wguide}
\end{center}
\end{figure}

As shown in Refs. \cite{Suresh0,Suresh},  a Hamiltonian of the form (\ref{H2}) can be used to describe a two-dimensional electron wave guide as seen in Fig. \ref{wguide}.

This wave guide can be  constructed by superposing two closed identical cavities and a lead. This forms the unperturbed system. The interaction appears as the cavities and the lead are connected. 

In suitable units the  horizontal dimension  of the cavities is $1$, and the vertical dimension  is $D$.  The lead has a horizontal dimension  $L\to  \infty$ and a vertical dimension $W$.  We consider a non-relativistic  electron, neglecting the spin.

If the electron is placed inside a closed cavity, its wave functions correspond to discrete cavity modes. The cavity modes can be labeled as 
$ |m, n\ket$, where $m$, $n$ are positive integers representing the  horizontal and vertical wave numbers.  The corresponding energies are
\begin{eqnarray}
 \xi^{m,n} = m^2 + n^2/D^2
   \EQN{Emn}
\end{eqnarray}

An electron placed inside the lead (with no cavities) has modes that can be labeled as $|k,l\ket$ where
$kL/\pi$ is the horizontal wave number and $l$ the vertical wave number.  The energies are
\begin{eqnarray}
  E_{k,l} = k^2/\pi^2 + l^2/W^2
 \EQN{Ekj}
\end{eqnarray}
As $L\to\infty$,   $k$ becomes a continuous variable.  On the other hand, $m$, $n$ and $l$ are always integers.

We consider an electron with low energy narrowly centered around
\begin{eqnarray}
 \xi^0 = \xi^{m_0,n_0}
    \EQN{xi0}
\end{eqnarray}
We assume that
\begin{eqnarray}
 E_{0, 1}< \xi^0 < E_{k, l}
    \EQN{xib}
\end{eqnarray}
for $l>1$ and all $k$.  The electron may propagate through the first mode of the lead, but not through the $l>1$ modes.

We also assume that there are no other cavity modes with energy between $E_{0, 1}$ and $\xi_0$.
Under these conditions, the cavity mode $\xi^0$ behaves essentially like the excited state in the Friedrichs-Lee model. It will decay with a finite lifetime. This means than an electron inside the cavity will escape through the lead.

The   following approximate Hamiltonian is obtained \cite{Suresh}.
\begin{eqnarray}
  H_{WG} &=& \xi^0[ |1\ket \bra 1| +   |2\ket \bra 2| ] +  \sum_{k,l} E_{k,l} |\psi_{k,l}\ket \bra \psi_{k,l}| \nonumber\\
  &+& \sum_{k,l}  V^0_{k,l}  \left( |1\ket \bra \psi_{k,l}| e^{ik x_1} 
   +  |2\ket \bra \psi_{k,l}| e^{ik x_2}  \right)
   + \rm{h.c.} \nonumber\\
 \EQN{HWG}
\end{eqnarray}
Here, h.c. means Hermitian conjugate. The cavities are centered at $x=x_1$ and $x=x_2$, where $x$ is the horizontal coordinate. The states $|i\ket$ represent the electron inside cavity $i=1$ or $2$, occupying the  mode $\xi_0$. The states $| \psi_{k,l}\ket$ are modified lead modes; they essentially represent the electron inside the part of the lead that does not overlap with the cavities. The terms $V^0_{k,l}$ represent the amplitude of a transition of the electron from this part of the lead to the cavities or vice versa. Their detailed expression is given in Ref. \cite{Suresh}.

Except for the additional index $l$ and the dispersion relation (\ref{Ekj}), which is different from $\ome_k = |k|$, the Hamiltonian $H_{WG}$ is the same as our two-atom Hamiltonian (\ref{H2}).  As the two cavities are identical, we have a system analogous to the two identical atoms. 
Since there is only one continuous variable $k$ describing the propagation along the lead, we can think of the wave guide system as a one-dimensional system, with internal degree of freedom $l$ (note that $l$ is discrete). 

Using the results of the Sec. \ref{sec:two} we obtain the equation for the complex energy of the collective 
 state
 \beqa
  z^0_j = \xi^0 + 2 \int_0^\infty dk \sum_{l=1}^\infty \frac{ |v^0_{k,l} |^2}{(z^0_j -E_{k,l})^+}
  (1 + \sigma_j \cos k x_{21}) \nonumber\\
  \EQN{z0j}
  \eeqa
 where $v^0_{k,l} = (L/\pi) V^0_{k,l}$. 
We will show that there is a solution with vanishing decay rate, corresponding to  a stable collective state. We follow the procedure shown in Appendix \ref{app:gamma}. For a vanishing  decay rate we write $z^0_j = \xit^0_j - i\eps$, where $\eps>0$ is infinitesimal. This gives the following condition
on $x_{21}$:
 \beqa
  1 + \sigma_j \cos k_0 x_{21} = 0
  \EQN{WGg}
  \eeqa
where $k_0$ is a wave vector that satisfies 
 \beqa
  E_{k_0,l} = \xit^0_j, \quad l=1
  \EQN{WGg2}
  \eeqa
 Through these two equations, $x_{21}$ becomes a function of $\xit^0_j$, 
 \beqa
  x_{21} =  g(\xit^0_j) \equiv \frac{n}{\sqrt{\xit^0_j - E_{0,1}}}
  \EQN{WGg2'}
  \eeqa
 where $n=$ odd integer for $j=s$, and $n=$ even integer for $j=a$.
 The renormalized energy $ \xit^0_j$  is then given by the solution of the integral equation
 \beqa
\xit_j^0 &=& \xi_0 \EQN{WGg3}\\
              &+&2 \int_0^\infty dk \sum_{l=1}^\infty   |v_{k,l}^0|^2 
                  \frac{{\cal P}}{\xit_j^0 -E_{k,l} }  \left[1 + \sigma_j \cos[ k g(\xit^0_j) ] \right]  
                 \nonumber
  \eeqa
where ${\cal P}$ means principal part.  Similar to \Ep{G8}, this equation has a solution if the condition
 \beqa
\xi^0 - E_{0,1} > 2 \int_0^\infty dk \sum_{l=1}^\infty   |v_{k,l}^0|^2 
                  \frac{{\cal P}}{E_{k,l} - E_{0,1}} 
                \EQN{exist}
  \eeqa
is satisfied. If the two cavities are not too close,  we can replace the interaction $v_{k,l}^0$ by the interaction in a single-cavity system. Then, \Ep{exist}  is essentially the condition that the electron in an individual cavity  has enough energy $\xi^0$ to escape through the lead. This condition is analogous
to \Ep{w1u}.

 In summary, adjusting the distance between the cavities, so that \Ep{WGg2'} is satisfied, 
we obtain a collective stable state where the electron remains trapped inside the two cavities, in either a symmetric or antisymmetric state.  The electron is trapped even though it would escape if there was only one cavity. 

To obtain this result we neglected the influence of cavity modes other than $\xi^0$. The existence of stable configurations in the wave guide could be verified by other methods, including numerical simulations or experiments. 

\section{Concluding remarks}

We have analyzed a two-atom system using complex collective  eigenstates of the Hamiltonian. 
Our main result is the description of long-range effects in one-dimensional space, such as the vanishing of the decay rate at regular intervals of distance $x_{21}$.  Another result is the application of  this model to two-dimensional electron wave guides, where a two-cavity configuration can be tuned to act as an electron trap. 

In our two-atom  model we neglected virtual transitions. This corresponds to a rotating wave approximation. Phenomena such as the existence of collective stable states in one-dimensional atoms deserve further study, with the inclusion of virtual transitions. Electron wave guides, on the other hand, are already well described by the type of Hamiltonian we considered, without virtual transitions. The description improves if we include more  cavity modes \cite{Suresh}.

Emitted photons are described by an exponentially growing field, truncated at the light cone of the atoms. This field plays an important role in the two-atom system, giving a strong influence on the lifetime or average energy of the collective states.  This field is directly related to the exponential decay of unstable states, which can be regarded as one of the simplest dissipative phenomena on a microscopic scale. So, in a sense, the formation of collective states is a microscopic non-equilibrium process, driven by  dissipation.

 \acknowledgments     %ACKNOWLEDGMENTS

We   thank   Professors R. Passante, T. Petrosky,  L. Reichl,  and W. Schieve, as well as Dr. G. Akguc, R. Barbosa, Dr.  E. Karpov, Dr.   C.B. Li,  A. Shaji,  and  M. Snyder for  helpful comments and suggestions.  We acknowledge the International Solvay Institutes for Physics and Chemistry, the Engineering Research Program of the Office of Basic Energy
Sciences at  the U.S. Department of Energy, Grant No DE-FG03-94ER14465, the
Robert A. Welch Foundation Grant F-0365, and the European Commission Project HPHA-CT-2001-40002 for supporting  this work.

%%%%%%%%%%%%%%%%%%%%
% Appendices      %
%%%%%%%%%%%%%%%%%%%%

 \appendix

%%%%%%%%%%%%%%%%%%%%%%%%%%
\section{Proof of \Ep{bces6}}
\label{app:zszs1}
%%%%%%%%%%%%%%%%%%%%%%%%%%%
 We start with the expression (see \Ep{bces3})
\beqa
  {\It}_0(t) =  \frac{1}{2\pi i} \sum_{n=0}^{\infty} \int_C dk\,
 \frac{\Del(k)^n }
 {(k-z_{s1})^{n+1}}e^{-i k t}
  \EQN{bces7}
\eeqa
where $C$ is a clockwise contour surrounding $k=z_{s1}$. Taking the residues at this point we get
\beqa
  {\It}_0(t) = \sum_{n=0}^{\infty}
 \frac{-1}{n!} \frac{\pard^n}{\pard k^n} [\Del(k)^n e^{-i k t} ]_{k=z_{s1}} 
  \EQN{bces8}
\eeqa
This is a perturbation expansion around $z_{s1}$, so it will correspond only the contribution from the pole $z_s$ and not
the poles $z_{s,n}$. 

We will show that 
\beqa
  \frac{\pard}{\pard t} {\It}_0(t) = - i z_s {\It}_0(t).
  \EQN{bces9}
\eeqa
Together with \Ep{bces4} this will  prove \Ep{bces6}. 
Starting with \Ep{bces8} and using
\beqa
  \frac{\pard^n}{\pard k^n} AB =  \sum_{l=0}^n \frac{n!}{l!(n-l)!} \left[  \frac{\pard^{n-l}}{\pard k^{n-l}} A \right]  \left[  \frac{\pard^l}{\pard k^l} B \right]
  \EQN{chain}
\eeqa
we obtain
\beqa
  \frac{\pard}{\pard t} {\It}_0(t) = - i z_{s1} {\It}_0(t) - i \It_1(t)
  \EQN{bces10}
\eeqa
where 
\beqa
  {\It}_m (t) =  \sum_{n=0}^{\infty}
 \frac{-1}{n!} \frac{\pard^n}{\pard k^n} [\Del(k)^{n+m}  e^{-i k t}]_{k=z_{s1}}
  \EQN{bces11}
\eeqa
Using \Ep{chain} again we get
\beqa
  {\It}_m (t) =  \sum_{l=0}^{\infty}
 \frac{1}{l!} \It_l(t) \frac{\pard^l}{\pard k^l}  [\Del(k)^{m}]_{k=z_{s1}}  
  \EQN{bces12}
\eeqa
With   \Ep{Delk4} for $z_{s,0} =  z_s$  and the Taylor expansion of $\Del(z_s)^l$ around $z_{s1}$ we find that the solution of this system of equations is
\beqa
  {\It}_l (t) =  \Del(z_s)^l \It_0(t)
  \EQN{bces13}
\eeqa
which combined with \Ep{bces10} proves \Ep{bces9}.

%%%%%%%%%%%%%%%%%%%%%%%%%%%%%%%
\section{Oscillations of $\gamma_j$ and $\ome_j$ with $x_{21}$}
\label{app:gamma}
%%%%%%%%%%%%%%%%%%%%%%%%%%%%%%%%

In this Appendix we will show that the decay rates $\gamma_j$ and energies $\omet_j$ of the collective states $|\phi_j\ket$  oscillate with the distance $x_{21}$ between the atoms. First we will  show that $\gamma_s$ vanishes (comes infinitesimally close to zero)
for distances 
\beqa
[x_{21}]_{\gamma_s=0} = \frac{(2n+1) \pi}{\omet_s^o} 
 \EQN{G1}
 \eeqa
 where $n$ is an integer, and 
 \beqa
 \omet_s^o = [\omet_s]_{\gamma_s=0} 
 \EQN{ometo}
 \eeqa
Similarly we  will show that decay rate $\gamma_a$ vanishes for
\beqa
 [x_{21}]_{\gamma_a=0} = \frac{2n \pi}{\omet_a^o}
 \EQN{G2}
 \eeqa
 where 
 \beqa
 \omet_a^o = [\omet_a]_{\gamma_a=0}. 
 \EQN{ometa}
 \eeqa
We start with the equation $\eta^+_s(z_s)=0$ or 
 \beqa
  z_s = \ome_1 + 2 \int_0^\infty dk \frac{\lam^2 v_k^2}{(z_s -k)^+}(1 + \cos k x_{21})
  \EQN{G3}
  \eeqa
 Assuming $z_s=\omet_s^o - i\eps$ with infinitesimal $\eps$ we have
 \beqa
  \omet_s^o &=& \ome_1 + 2 \int_0^\infty dk \frac{\lam^2 v_k^2}{\omet_s^o -k + i\eps}(1 + \cos k x_{21})
                           \nonumber\\
                  &=& \ome_1 + 2 \int_0^\infty dk \lam^2 v_k^2 \left[
                  \frac{{\cal P}}{\omet_s^o -k } - \pi i \del(\omet_s^o-k)\right] \nonumber\\
                  &\times& (1 + \cos k x_{21})
  \EQN{G4}
  \eeqa
 where  we used the relation
 \beqa
 \frac{1}{\ome + i\eps} = \frac{ {\cal P}}{\ome} - \pi i \del(\ome)
 \EQN{Pdel}
 \eeqa
 together with \Ep{complexd}. 
  Comparing the left and right-hand sides of \Ep{G4} we see that  the imaginary part should vanish, so we get
 \beqa
  1 + \cos \omet_s^o x_{21} = 0
  \EQN{G5}
  \eeqa
 which proves \Ep{G1}.
 In a similar way, starting from the equation for $z_a$, 
 \beqa
  z_a = \ome_1 + 2 \int_0^\infty dk \frac{\lam^2 v_k^2}{(z_a -k)^+}(1 - \cos k x_{21})
  \EQN{G6}
  \eeqa
 we get
 \beqa
  1 - \cos \omet_a^o x_{21} = 0
  \EQN{G7}
  \eeqa
 which proves \Ep{G2}. 
  
 The $\omet_j^o$ satisfy the integral equations
 \beqa
  \omet_s^o &=& \ome_1 + 2 \int_0^\infty dk \lam^2 v_k^2 
                  \frac{{\cal P}}{\omet_s^o -k }  \left[1 + \cos \frac{(2n+1)\pi k}{\omet_s^o}\right] \nonumber\\
  \omet_a^o &=& \ome_1 + 2 \int_0^\infty dk \lam^2 v_k^2 
                  \frac{{\cal P}}{\omet_a^o -k }  \left[1 - \cos \frac{2n\pi k}{\omet_a^o}\right] \nonumber\\
  \EQN{G8}
  \eeqa
 Using graphical methods it can be shown the first equation has a unique solution for each integer $n$, provided that  
 \beqa
   \ome_1 - 2 \lim_{\omet_s^o\to 0} \int_0^\infty dk \lam^2 v_k^2 
                  \frac{{\cal P}}{k }  \left[1 + \cos \frac{(2n+1)\pi k}{\omet_s^o}\right] > 0\nonumber\\
                    \EQN{G8'}
  \eeqa
  In  the limit $\omet_s^o\to 0$ the cosine term gives a vanishing integration. Thus \Ep{G8'}  is satisfied if \Ep{w1u} is satisfied.  A similar argument applies to the second equation in (\ref{G8}). 
  
 Eqs. (\ref{G1}) and  (\ref{G2}) explain the oscillatory behavior of $\gam_j$ seen  in Fig. \ref{gams}. 

To explain the oscillations of $\omet_j$ we note that the terms inside brackets in \Ep{G8} are even in $k$ around $\omet_j^o$, regardless of $n$. 
On the other hand, the principal parts are odd. Hence the product is odd and the integration around $\omet_j^o$ vanishes. Thus the largest contributions to the integrals come from the tails of the principal parts. 
The ``$1$'' terms inside the brackets give a much larger contribution than the ``$\cos$'' terms, because the latter oscillate with $k$. Neglecting the ``$\cos$'' terms we get
 \beqa
 \omet_s^o \approx \omet_a^o \approx \omet_1
   \EQN{G9}
  \eeqa
where $\omet_1$ is the one-atom shifted energy (see \Ep{SUM2'}).  This shows that the $\omet_j$ have approximately the same values when their respective $\gamma_j$ vanish. From \Ep{G9} we conclude that the period of the oscillations of $\gamma_j$ and $\omet_j$ is approximately $2\pi/\omet_1$.

Adding Eqs. (\ref{G3}) and  (\ref{G6}) we see that for weak coupling the poles of the one and two-atom Green functions obey the relations
 \beqa
 z_1 \approx \frac{z_a+z_s}{2}
   \EQN{G13}
  \eeqa
So both $\omet_j$ and $\gamma_j$ oscillate around the one-atom $\omet_1$ and $\gamma_1$, respectively.

Finally, we show that the ``force'' ${\cal F}_s$ between the atoms is a maximum when the decay rate is zero, as seen in Fig. \ref{Fng}. When $\gamma_s=0$, we have
 \beqa
& &\frac{d{\cal F}_s^o}{dx_{21}} =  -\frac{d^2\omet_s^o}{dx_{21}^2} \nonumber\\
&\approx& 2 \int_0^\infty dk \lam^2 v_k^2 
                 \frac{ {\cal P}}{\omet_s^o -k }  k^2 \cos \frac{(2n+1)\pi k}{\omet_s^o}
   \EQN{G14}
  \eeqa
 As argued above \Ep{G9} the integral of the cosine  is small. Hence we have
 \beqa
\frac{d{\cal F}_s^o}{dx_{21}} \approx 0
   \EQN{G15}
  \eeqa
 A similar argument may be applied to  ${\cal F}_a$.

%%%%%%%%%%%%%%%%%%%%%%%%%%%%%%%
\section{Sub-radiance in $d>1$ dimensions}
\label{app:23}

In one dimension, the vanishing of the collective decay rate occurs for distances given by the conditions 
 \beqa
  1 + \sigma_j \cos \omet_j^o x_{21} = 0
  \EQN{dd1}
  \eeqa
Assuming the potential $v_k$ is rotationally invariant, in $d>1$ dimensions, analogous conditions would be
 \beqa
  \int_{0}^{\pi}  \Ome(\theta) d\theta [1 + \sigma_j \cos (\omet_j^o x_{21} \cos\theta)] = 0
  \EQN{dd2}
  \eeqa
where $\theta$ is the angle of  the wave  vector ${\bf k}$, with respect to the line joining the two atoms. The function $\Ome$ is $2$ for $d=1$ and $\sin\theta$ for $d=3$. 

We see that \Ep{dd2} can only be satisfied for the antisymmetric state with $\sigma_j=-1$ and for short distances  $x_{21} \ll \omet_j^{o} \approx \omet_1^{-1}$. This agrees with the results of Stephen \cite{Stephen} anticipated by Dicke \cite{Dicke}.

%%%%%%%%%%%%%%%%%%%%
% References       %
%%%%%%%%%%%%%%%%%%%%

%%%%%%%%%%
% End    %
%%%%%%%%%%

\end{document}